\documentclass[twocolumn]{aastex62}

\usepackage[utf8]{inputenc}
\usepackage[T1]{fontenc}
\usepackage{wasysym}
\usepackage{bm}
\usepackage{multirow}

\newcommand{\Msun}{~\text{M}_{\astrosun}}
\newcommand{\E}[1]{\times 10^{#1}}

\graphicspath{{./}{figures/}}

\received{2020, April 7}
\revised{2020, May 8}
\accepted{2020, May 31}
\submitjournal{ApJL}

\shorttitle{Large-scale turbulent driving regulates star formation}

\begin{document}

\title{Large-scale turbulent driving regulates star formation in high-redshift gas-rich galaxies}

\author{Noé Brucy}
\affiliation{AIM, CEA, CNRS, \\
Université Paris-Saclay, Université Paris Diderot, Sorbonne Paris Cité \\
F-91191 Gif-sur-Yvette, France}

\author{Patrick Hennebelle}
\affiliation{AIM, CEA, CNRS, \\
Université Paris-Saclay, Université Paris Diderot, Sorbonne Paris Cité \\
F-91191 Gif-sur-Yvette, France}

\author{Frédéric Bournaud}
\affiliation{AIM, CEA, CNRS, \\
Université Paris-Saclay, Université Paris Diderot, Sorbonne Paris Cité \\
F-91191 Gif-sur-Yvette, France}

\author{Cédric Colling}
\affiliation{AIM, CEA, CNRS, \\
Université Paris-Saclay, Université Paris Diderot, Sorbonne Paris Cité \\
F-91191 Gif-sur-Yvette, France}

\nocollaboration



\begin{abstract}

The question of what regulates star formation is a longstanding issue. To investigate this issue, we run simulations of a kiloparsec cube section of a galaxy with three kinds of stellar feedback: the formation of H~II regions, the explosion of supernovae, and the ultraviolet heating.
We show that stellar feedback is sufficient to reduce the averaged star formation rate (SFR) to the level of the Schmidt-Kennicutt law in Milky Way-like galaxies but not in high-redshift gas-rich galaxies suggesting that another type of support should be added. 
We investigate whether an external driving of the turbulence such as the one created by the large galactic scales could diminish the SFR at the observed level.
Assuming that the Toomre parameter is close to 1 as suggested by the observations, we infer a typical turbulent forcing that we argue 
should be applied parallel to the plane of the galactic disk. When this forcing is applied in our simulations, the SFR within our simulations closely follows the Schmidt--Kennicutt relation. We found that the velocity dispersion is strongly anisotropic with the velocity dispersion alongside the galactic plane being up to 10 times larger than the perpendicular velocity.

\end{abstract}



\section{Introduction}\label{sec:intro}

The formation of stars is a key process that has a major impact on the galactic evolution.
Its efficiency and rate are influenced by many factors, and the relative importance of each of them is still poorly understood.
One of the main reasons why it is so hard to fully understand star formation is that it involves scales ranging from a few astronomical units up to several kiloparsecs, with about nine orders of magnitude between them. As a consequence, self-consistent simulations of star formation in a galaxy are out of reach for now, and some possibly important factors have to be neglected or added through subgrid models \citep{duboisOnsetGalacticWinds2008,hopkinsSelfregulatedStarFormation2011}. 
Simulations of smaller regions of a galaxy are a useful complementary tool that enables the use of a higher resolution and the performance of parametric studies.
An important challenge for this kind of numerical simulations is reproducing the Schmidt--Kennicutt (SK) law \citep{kennicuttjr.GlobalSchmidtLaw1998,kennicuttStarFormationMilky2012} that links the star formation rate (SFR) to the column density of gas.
Previous results \citep{walchSILCCSImulatingLifeCycle2015,padoanSupernovaDrivingOrigin2016,iffrigStructureDistributionTurbulence2017,kimThreephaseInterstellarMedium2017,gattoSILCCProjectIII2017} indicated that the magnetic field has a moderate effect on the SFR but that stellar feedback (namely H~II regions and supernovae) can greatly reduce the SFR in Milky Way-like galaxies down to a rate consistent with the observed one.
\cite{collingImpactGalacticShear2018} have shown that with a more comprehensive model of stellar feedback,  including ionizing radiation as well as supernovae that explode after a delay corresponding to the stellar lifetime, the SFR typically lies a few times above the SK relation.
However, they have shown that the galactic shear may be able, if it is strong enough, to reduce the SFR sufficiently to make it compatible with the SK law. 
In our work, we run simulations of a local region of a galactic disk within a kiloparsec cube box. We use a numerical setup that is very close to the one used by \cite{collingImpactGalacticShear2018}. Our primary goal is to extend their results to galaxies with higher column densities with the aim to reproduce the SK law. The galaxies that we model have a stellar and dark matter potential similar to the Milky Way with a mean column density of gas $\Sigma_{0,\mathrm{gas}}$ that varies from $13$ to $155 \Msun\cdot\mathrm{pc}^{-2}$, representative for Milky Way-like galaxies up to gas-rich galaxies at redshift $z =$~1--3 \citep{genzelRingsBulgesEvidence2008,genzelStudyGasstarFormation2010,daddiVeryHighGas2010}. 
Since the total gravitational potential remains constant, so does the galactic shear, which is therefore not sufficient to regulate star formation  \citep{collingImpactGalacticShear2018}.
On the other hand, several recent studies have shown that injection of turbulence from galactic motions has to be taken into account in order to explain the observed velocity dispersion and SFR \citep{renaudStarFormationLaws2012,krumholzUnifiedModelGalactic2018,meidtModelOnsetSelfgravitation2020} as suggested by \cite{bournaudISMPropertiesHydrodynamic2010}. 
Possible source of turbulence include the orbital energy or even mass accretion onto the galaxies. The latter in particular requires a mechanism such as an instability to degrade this source of free energy.  We test the effect of such injection of turbulence by adding a large-scale turbulent driving similar to the one used by \cite{schmidtNumericalSimulationsCompressively2009}. 

This Letter is organized as follows. In the section \ref{sec:setup} we present our numerical setup and our simulations.
In section \ref{sec:sf} we investigate the relation between the SFR and the gas column density when only stellar feedback is at play. In section \ref{sec:turb} we show the results of similar simulations when we add a turbulent driving. 
The necessity of the stellar feedback to quench star formation is investigated in section \ref{sec:full_turb}. Section \ref{sec:conc} concludes the Letter.

\section{Numerical setup}\label{sec:setup}

\begin{table}
\caption{List of Simulations.}\label{tbl:simu}
  \begin{center}
	\begin{tabular}{llcccc}
	\hline\noalign{\smallskip}
	Group & $n_0$ & $f_{\text{rms}}$ & $\Sigma_{0, \mathrm{gas}}$ & $\overline{P_{\mathrm{inj}}}$ \\
	&{\scriptsize $[\mathrm{cm}^{-3}]$ }& &{ \scriptsize $[\Msun\cdot\mathrm{pc}^{-2}]$}&{\scriptsize  $[\mathrm{W}]$ }\\
	\noalign{\smallskip}
	\hline
	\noalign{\smallskip}
	\hline
	\noalign{\smallskip}
	\multirow{7}{*}{\textsc{noturb}} 
	& 1 & 0 & 12.9 & 0  \\
	& 1.5 & 0 & 19.4 & 0 \\
	&  2 & 0 & 25.8 & 0 \\
	&  3 & 0 & 38.7 & 0 \\
	&  4 & 0 & 51.6 & 0 \\
	&  6 & 0 & 77.4 & 0 \\  
	&  12 & 0 & 155 & 0 \\  
	\hline
	\multirow{4}{*}{\textsc{turb2.5}} 
	& 1.5 & $2.5 \E{4}$ & 19.4 & $1.7 \pm 0.7 \E{31}$\\
	&  3 & $6.0 \E{4}$ & 38.7 & $9.1 \pm 3.8 \E{31}$\\ 
	&  6 & $1.0 \E{5}$ & 77.4 & $5.6 \pm 3.6 \E{32}$\\
	 &  12 & $2.0 \E{5}$ & 155 & $3.1 \pm 2.2 \E{33}$\\
	\hline
	\multirow{4}{*}{\textsc{turb3.8}}
	& 1.5 & $2.0 \E{4}$ & 19.4 & $1.1  \pm 0.5 \E{31} $ \\
	&  3 & $8.0 \E{4}$ & 38.7 & $1.7 \pm 1.1 \E{32} $\\
	&  6 & $2.0 \E{5}$ & 77.4 & $1.6 \pm 1.3 \E{33}$\\
	&  12 & $1.0\E{6}$ & 155 & $3.4 \pm 3.0 \E{34}$ \\  
	\hline
	\multirow{4}{*}{\textsc{nofeed}}
	& 1.5 & $2.0 \E{4}$ & 19.4 & $1.1  \pm 0.5 \E{31} $ \\
	&  6 & $2.0 \E{5}$ & 77.4 & $1.6 \pm 1.3 \E{33}$\\
	&  12 & $1.0\E{6}$ & 155 & $3.4 \pm 3.0 \E{34}$ \\  
	\noalign{\smallskip}
	\hline
	\multicolumn{6}{p{8.5cm}}{\textbf{Note.} The total averaged injected power $\overline{P_{\mathrm{inj}}}$ is computed by comparing the kinetic energy in the box before and after applying the turbulent force. \added{Simulations in the \textsc{nofeed} group has no stellar feedback (see section \ref{sec:full_turb}).}}
	\end{tabular} 
  \end{center}
\end{table}

\subsection{Magnetohydrodynamic (MHD) Simulations}\label{subsec:mhd}

We use the RAMSES code \citep{teyssierCosmologicalHydrodynamicsAdaptive2002}, to solve the equations of MHD with a Godunov solver \citep{fromangHighOrderGodunov2006} on a cubic grid of $256^3$ cells \added{with periodic boundaries on the midplane and open vertical boundaries}.
The box represents a cubic region of the galactic disk of size $L = 1$\,kpc, so the resolution is about 4\,pc. 
Sink particles \citep{bleulerMoreRealisticSink2014} are used to follow the dense gas and model star formation. Sink creation is triggered when the gas density overpasses a threshold of $10^3 \ \mathrm{cm^{-3}}$ \citep{collingImpactGalacticShear2018}. All the mass accreted by a sink is considered as stellar mass.

We use the same initial conditions as \cite{collingImpactGalacticShear2018}. To sum up, the gas (atomic hydrogen) is initially distributed as a Gaussian along $z$-axis,
  \begin{equation}
n(z) = n_0 \exp \left( - \frac{1}{2} \left( \frac{z}{z_0}  \right)^2 \right),
  \end{equation}
with $n_0$ a free density parameter and $z_0 = 150\ \mathrm{pc}$.  
The column density of gas (hydrogen and helium), integrated along the z-axis (perpendicular to the disk) is then
\begin{equation}
  \Sigma_{\text{gas}, 0} = \sqrt{2\pi} m_p n_0 z_0 
  \end{equation}
where $m_p = 1.4 \times 1.66 \cdot 10^{-24}$ g is the mean mass per hydrogen atom. 
The initial temperature is chosen to be $8000 \ \mathrm{K}$ to match the typical value of the temperature of the warm neutral medium (WNM) phase of the Interstellar Medium (ISM).
An initial turbulent velocity field with a root mean square (RMS) dispersion of $5~\mathrm{km\cdot s}^{-1}$  and a Kolmogorov power spectrum with random phase \citep{kolmogorovLocalStructureTurbulence1941} is also added.
Finally, we add a Gaussian magnetic field, oriented along the $x-axis$,
  \begin{equation}
B_x(z) = B_0 \exp \left( - \frac{1}{2} \left( \frac{z}{z_0} \right)^2 \right),
  \end{equation}
with $B_0 = 4\ \mathrm{\mu G}$. The rotation of the galaxy is not modeled.

\subsection{Stellar feedback}

The simulations include models for the formation and expansion of H~II region, explosion of supernovae (SNe) and the far-ultraviolet (FUV) feedback. The H~II and SN feedback models are same as in \cite{collingImpactGalacticShear2018}.
As in \cite{collingImpactGalacticShear2018}, the FUV heating is uniform. However, it is not kept constant at the solar neighborhood value because young O-B star contribute significantly to the FUV emission. As a first approximation, the FUV heating effect can be considered to be proportional to the SFR \citep{ostrikerRegulationStarFormation2010}. The mean FUV density relative to the solar neighbourhood value $G_0^{\prime}$ can then be written as
\begin{equation}
\label{eq:guv}
    G_0^\prime = \frac{\Sigma_{\mathrm{SFR}}}{\Sigma_{\mathrm{SFR,}\astrosun}}
               = \frac{\Sigma_{\mathrm{SFR}}}{2.5\times10^{-9}~\mathrm{M}_{\astrosun}\cdot\mathrm{pc}^{-2}\cdot\mathrm{yr}^{-1}}
\end{equation}
In our model, $G_0^{\prime}$ has a minimal value of $1$ (as a background contribution) and follows the equation \ref{eq:guv} when the SFR increases. 

\subsection{Injection of Turbulence}
\label{subsec:turb_inj}

\cite{bournaudISMPropertiesHydrodynamic2010}, \cite{krumholzTurbulenceInterstellarMedium2016} and
\cite{krumholzUnifiedModelGalactic2018} show that for galaxies with high column densities or high SFRs, large-scale gravitational instabilities are the main source of turbulent energy and dominate over stellar feedback.
We investigate numerically the effect of this turbulent driving on star formation. We use a model for turbulent driving adapted from the generalization of the Ornstein--Uhlenbeck used and explained by several authors  \citep{eswaranExaminationForcingDirect1988,schmidtNumericalDissipationBottleneck2006,schmidtNumericalSimulationsCompressively2009,federrathComparingStatisticsInterstellar2010}. 
The driving is bidimensional (2D) because we consider disk-shaped galaxies and expect large-scale turbulence driving to act mainly within the disk plane. A numerical confirmation of the predominance of the 2D modes at large scale in global galactic simulations is given by \cite{bournaudISMPropertiesHydrodynamic2010} in Figure~7 in this article.

More precisely, the turbulent forcing is described by an external force density $\bm{f}$ that accelerates the fluid on large scales. The evolution of the Fourier modes of the acceleration field $\bm{\hat{f}}(\bm{k}, t)$ follows 
\begin{equation}
\label{eq:ed_fourier}
    d\bm{\hat{f}}(\bm{k}, t) = - \bm{\hat{f}}(\bm{k}, t)\frac{dt}{T} + F_0(\bm{k})\bm{P_\zeta}\left(\left(\begin{array}{c} k_x \\ k_y \\ 0\end{array}\right)\right)\cdot d\bm{W}_t
\end{equation}
In this stochastic differential equation, $dt$ is the timestep for integration and $T$ is the autocorrelation time scale. In our simulations, we  $T = 0.5$~Myr and $dt/T = 1/100$. Tests shows that choosing different values for $T$ does not significantly impact the simulations. \added{The Wiener process $\bm{W_t}$ and} the projection operator $\bm{P_\zeta}$ \replaced{is}{are} defined as in \cite{schmidtNumericalSimulationsCompressively2009}, $\zeta$ being the solenoidal fraction. In our runs, $\zeta = 0.75$, and as a consequence the turbulent driving is stronger for the solenoidal modes. This choice of $\zeta$ is motivated by the fact that more compressive drivings are prone to bolster star formation instead of reducing it. Furthermore, this choice is in agreement with the value of $\zeta = 0.78\pm0.14$ found by  \cite{jinEffectiveTurbulenceDriving2017} in their simulation of a Milky Way--like galaxy.
Note that we apply it to a projection of the wavenumber $\bm{k}$ in the disk plane instead of $\bm{k}$ itself, so that the resulting force will have no vertical component. 
The forcing field $\bm{f}(\bm{x}, t)$ is then computed from the Fourier transform

\begin{equation}
\label{eq:injection}
\bm{f}(\bm{x}, t) = f_{\mathrm{rms}} \times \int\bm{\hat{f}}(\bm{k}, t) e^{i\bm{k}\cdot x} d^3\bm{k}
\end{equation}
The parameter $f_{\mathrm{rms}}$ is directly linked to the power injected by the turbulent force into the simulation.

\subsection{Estimation of the Injected Power}
\label{subsec:estimation}
With general considerations we can get an idea of the power injected by large-scale turbulence.
The specific power $\epsilon$ injected by turbulence at a given scale $l$ can be related with the typical speed of the motions $v_l$ at that scale. This being true for each scale $l$, there is the following relation between $\epsilon$ and the velocity dispersion of the gas $\sigma_g$.
\begin{equation}
\epsilon \sim \frac{v_l^3}{l} \propto \sigma_g^3
\end{equation}
The disk is supposed to be at marginal stability, so that the Toomre parameter is $Q \sim 1$. The Toomre parameter can be estimated as follows:
\begin{equation}
\label{eq:Q}
Q = \frac{\sigma_g \kappa}{\pi \Sigma_g G} \propto \frac{\sigma_g \kappa}{\Sigma_g}
\end{equation}
where $\kappa$ is the epicyclic frequency (which does not depend on the gas column density $\Sigma_g$). Equation \ref{eq:Q} can be rewritten $\sigma_g \propto \Sigma_g$, a relation outlined in both observational and computational studies of high-redshift galaxies \citep{genzelStudyGasstarFormation2010,dekelFormationMassiveGalaxies2009,bournaudStarFormationLaws2014}.
This leads to the following estimation for the specific power
\begin{equation}
\epsilon \propto \Sigma_g^{3}.
\end{equation}
Therefore the total power injected by large-scale motions $P_{\mathrm{LS}}$ scales as
\begin{equation}
P_{\mathrm{LS}} \propto \Sigma_g^{4}.
\end{equation}
In the appendix \ref{sec:pinj_abs}, we provide a more detailed estimation of the absolute value of $P_{\mathrm{LS}}$.

\begin{figure*}[h]
    \centering
    \includegraphics[width=0.85 \textwidth]{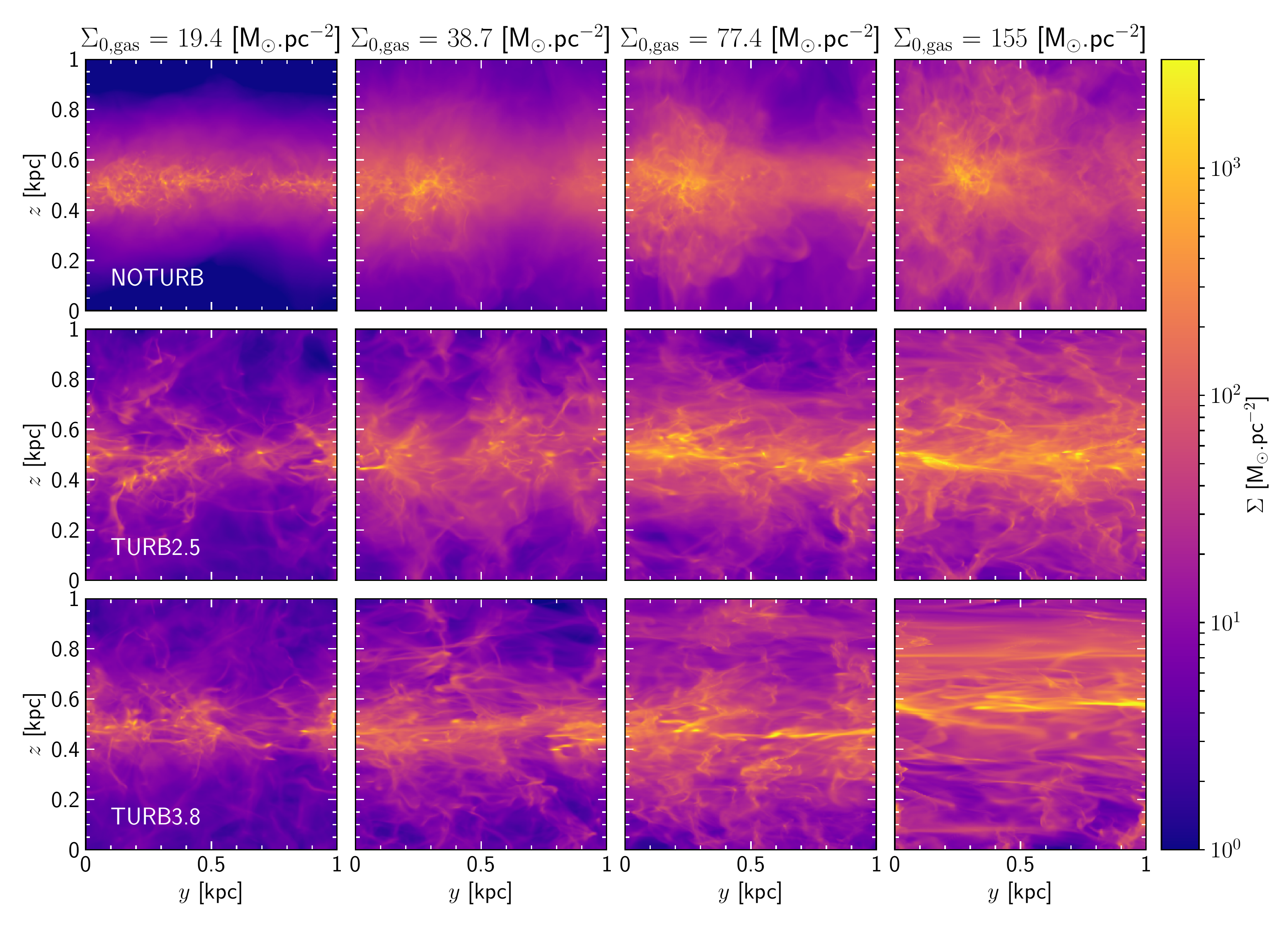}\\
    \includegraphics[width=0.85 \textwidth]{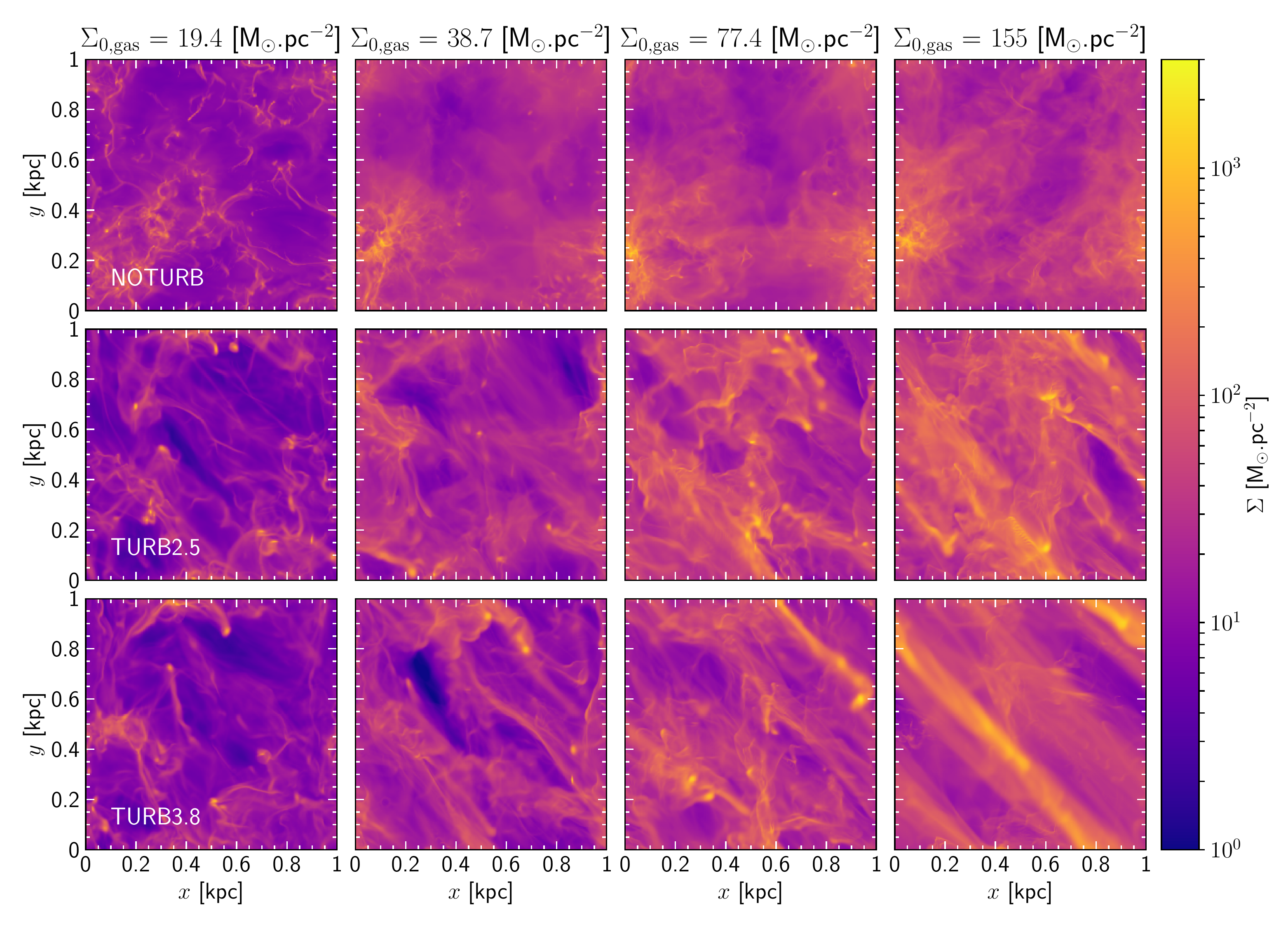}
    \caption{Column density maps, edge-on (top panel) and face-on (bottom panel). All snapshots are taken around $60$ Myr. The simulation without turbulence are dominated by the effects of the supernovae, while turbulent driving creates filamentary structures}\label{fig:coldens}
\end{figure*}

\begin{figure*}[ht!]
   \centering
   \includegraphics[width=0.49\textwidth]{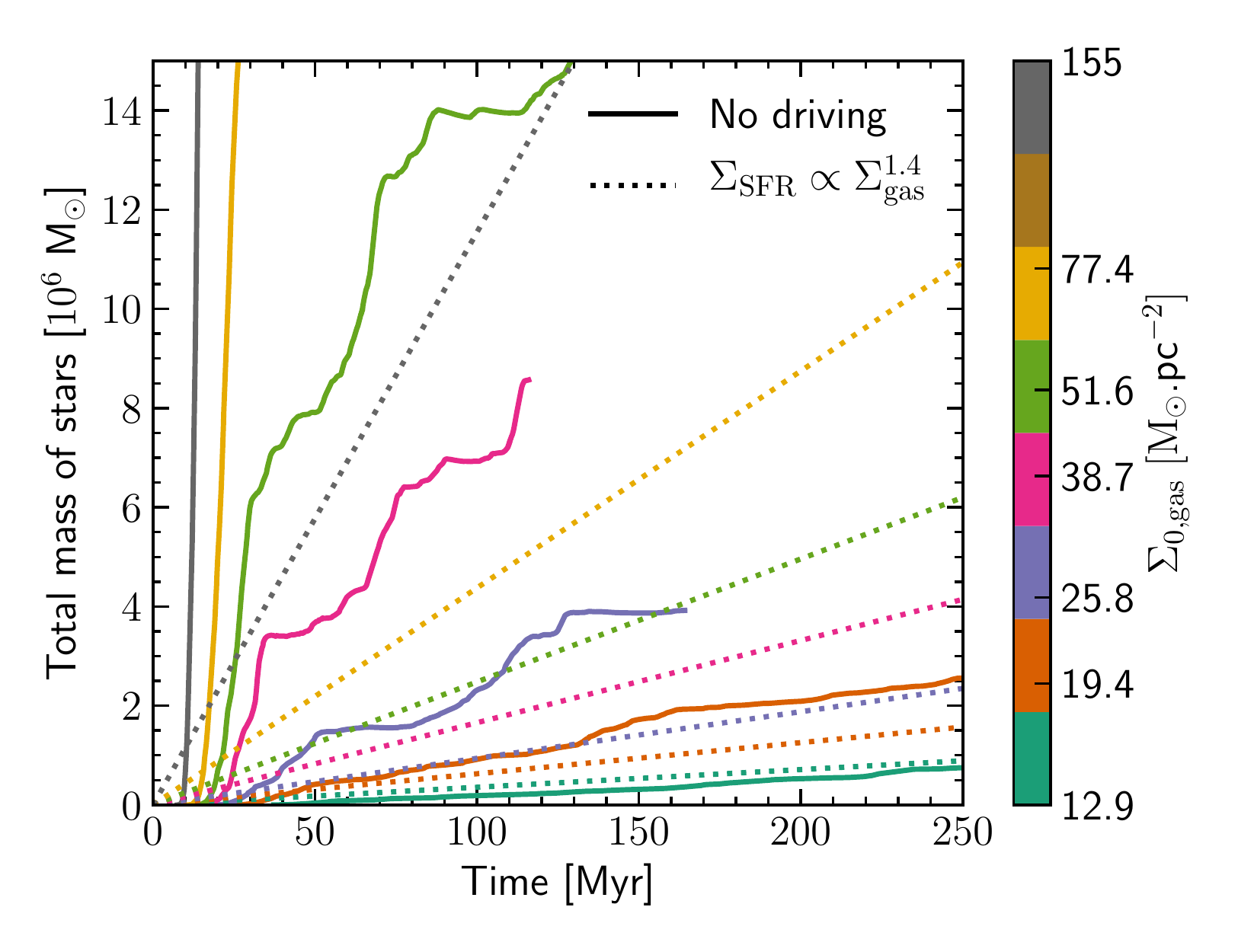}
   \includegraphics[width=0.49\textwidth]{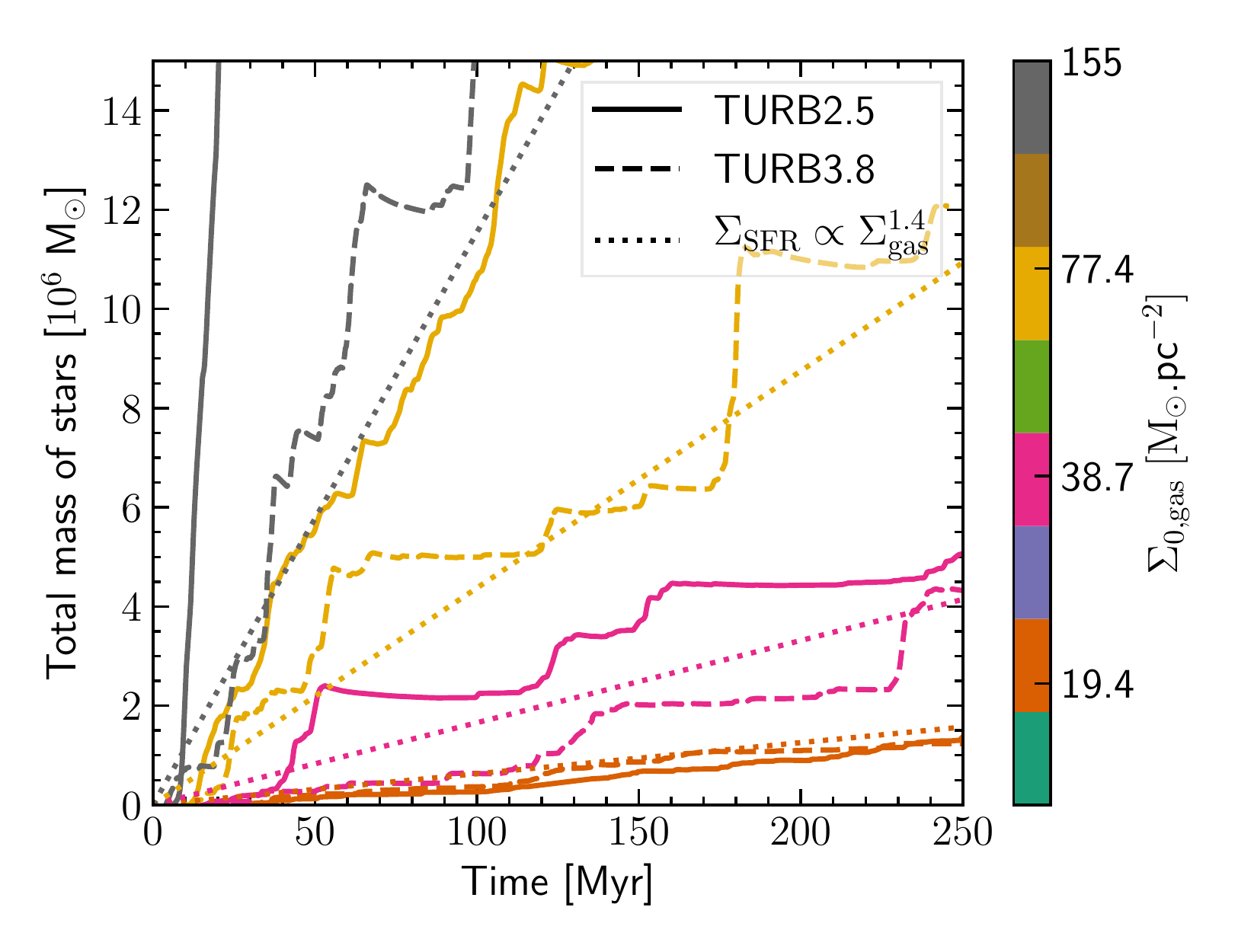}
    \caption{Evolution of the total stellar mass in the simulations. The total mass is compared to the stellar mass produced if the SFR was constant and matching the SK law (dotted lines). 
    With only the stellar feedback quenching the star formation, the star formation rate matches the Kennicutt law only for one simulation with $\Sigma_{\mathrm{0,gas}} = 12.9 ~\Msun\cdot\mathrm{pc}^{-2}$, slightly higher than the Milky Way. For higher column density, however, the SFR is well above the observed values. Adding the turbulent driving helps to reduce the SFR.}\label{fig:mass}
\end{figure*}

\subsection{List of Simulations}

In order to test the impact of the stellar feedback and the turbulent driving, we ran three groups of simulations. The list of the simulations is available in Table~\ref{tbl:simu}. Simulations within the group \textsc{noturb} have no turbulent driving and enable to test the efficiency of stellar feedbacks as star formation regulators.
In the group \textsc{turb2.5} the mean power injected $\overline{P_{\mathrm{inj}}}$ scales as $\Sigma_{0,\mathrm{gas}}^{2.5}$. The \textsc{turb3.8} has a stronger injection of turbulent energy, which scales as $\Sigma_{0,\mathrm{gas}}^{3.8}$, very close to $P_{\mathrm{LS}}$, the expected energy injected at large scales estimated in the section \ref{subsec:turb_inj} (see Figure \ref{fig:SK}b). Simulations in the \textsc{noturb} group have no stellar feedback.

\section{Pure Stellar Feedback Simulations}\label{sec:sf}

\begin{figure*}[ht]
\gridline{\fig{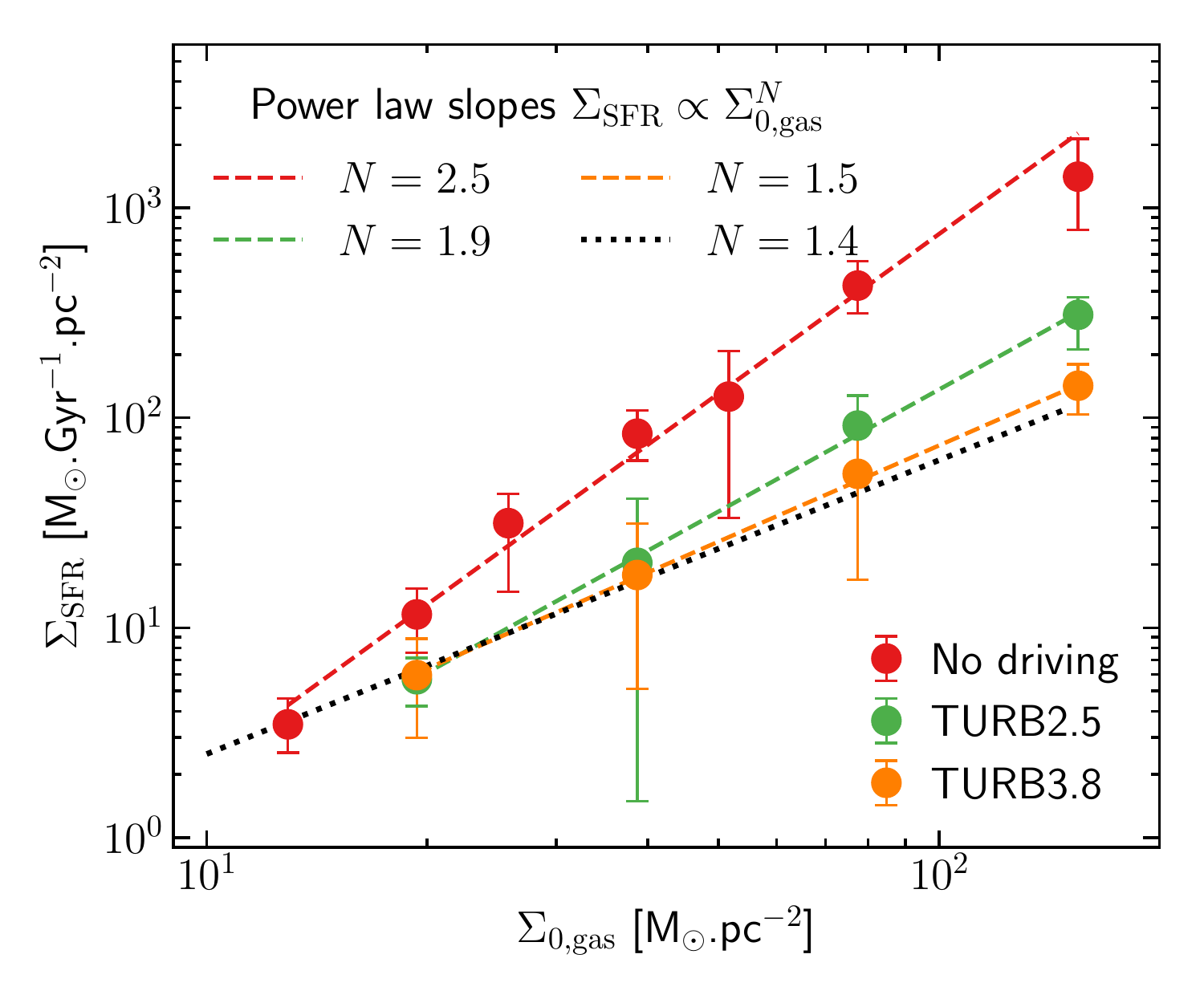}{0.49\textwidth}{(a)}
          \fig{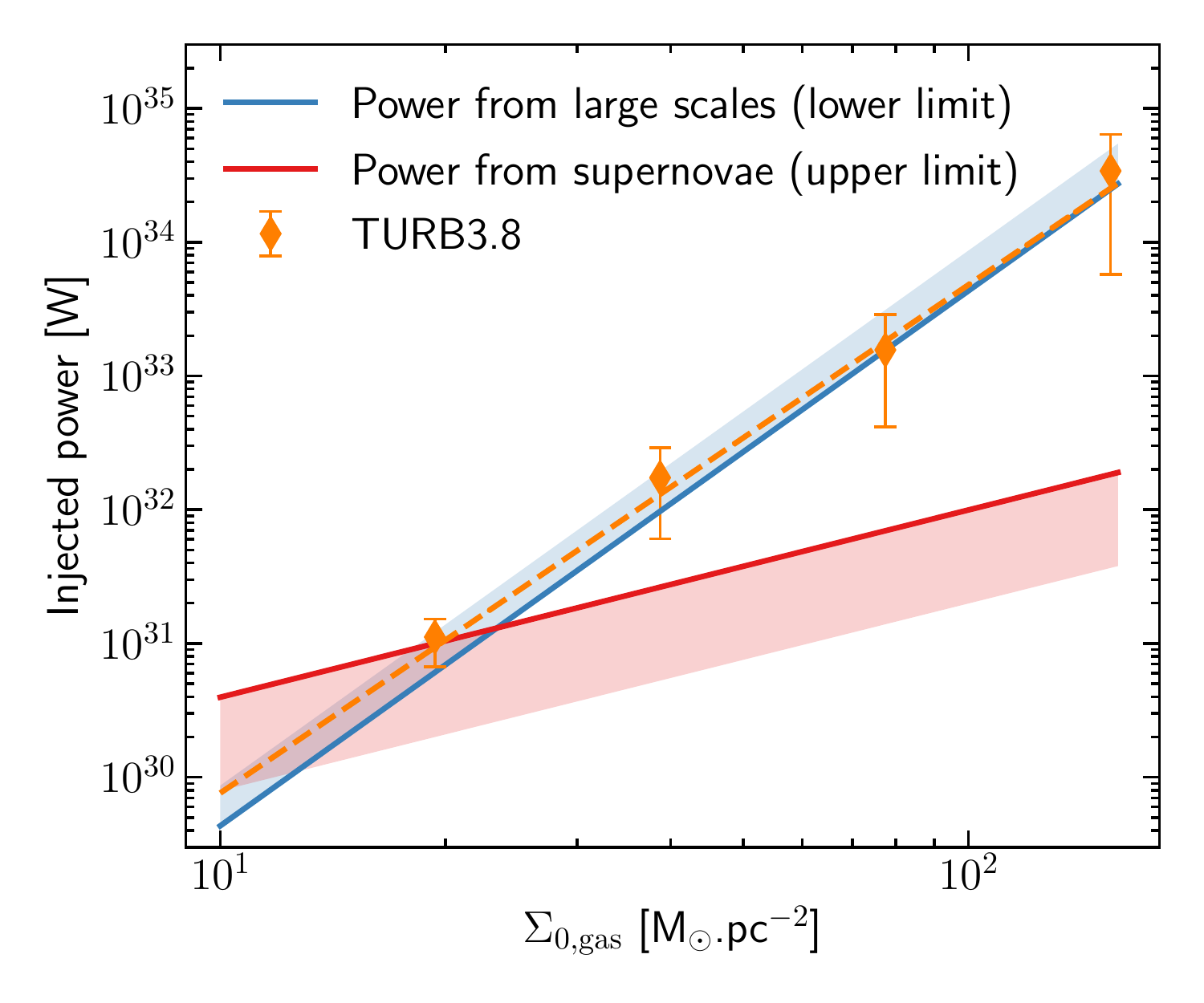}{0.49\textwidth}{(b)}}
    \caption{(a): Averaged surfacic SFR as a function of the initial column density. The SFR in computed at each step and averaged over a period of 40 Myr. 
    With pure stellar feedback the star formation law have an index of 2.5, and thus star formation is quenched enough only for the galaxies with moderate column density. With soft ($\overline{P_{\mathrm{inj}}} \propto \Sigma_{0, \mathrm{gas}}^{2.5}$) and strong $\overline{P_{\mathrm{inj}}} \propto \Sigma_{0, \mathrm{gas}}^{3.8}$) turbulent driving the obtained star formation is closer to the SK law, and even very close for the strong injection (with an index of 1.5).\\
(b): Injected power. The dotted orange line is fitted from our model \textsc{turb3.8} and is a power law of index 3.8 (see Table \ref{tbl:simu}). The blue and red filled lines are, respectively, an estimated lower bound for the turbulent power injected by large-scale motions ($P_{\mathrm{LS}}$) and an estimated upper bound for the power from the SNe converted into turbulence ($P_{\mathrm{SN}}$). \added{The shaded regions indicate a reasonable range for these values.} They are computed as explained in the appendix \ref{sec:pinj_abs}.}\label{fig:SK}
\end{figure*}

In this section we study the SFR when only stellar feedback regulates star formation (without additional turbulent driving, group \textsc{noturb}). Figure \ref{fig:coldens} features edge-on and face-on column density maps of the simulations. 
In \textsc{noturb} simulations, the gas tends to form clumpy structures. \replaced{The effects of the SNe explosions are clearly visible, especially}{Ejection of gas out of the disk plane due to supernovae explosions is clearly visible} in the simulations with a high initial gas column density.
Figure \ref{fig:mass} shows the evolution of the total sink mass during the simulation for several initial column density going from $\Sigma_{\text{gas}, 0}=12.9\Msun\cdot\text{pc}^{-2}$ to $\Sigma_{\text{gas}, 0}=155\Msun\cdot\text{pc}^{-2}$. 
The dotted lines correspond to the expected stellar mass growth if the SFR was constant and scaled as in the SK law. For $\Sigma_{\text{gas}, 0}=12.9\Msun\cdot\text{pc}^{-2}$ (corresponding to a galaxy slightly heavier than the Milky way) the SFR is close to the observed one for similar galaxies. \added{That means that for such galaxies, the feedback is strong enough to regulate the star formation rate. This is not true in the inner regions where the column density is higher and where the bar plays a considerable role in triggering and/or quenching star formation \citep{emsellemInterplayGalacticBar2015}, and in the outer regions without stars, but these regions represent a small fraction of the total mass of the galaxy.}
However, the stellar mass growth is considerably faster than expected in heavier galaxies, with SFR that can overpass the observation by more than one order of magnitude.
Interestingly, the SFR also follows a star formation law $\Sigma_{\mathrm{SFR}} \propto \Sigma_{\mathrm{gas}}^N$ (see Figure \ref{fig:SK}a), but with an index $N=2.5$, which is much steeper than the $N=1.4$ determined by Kennicutt.
This is unlikely to be due to an underestimation of the stellar feedback. First, all the main processes that may quench the star formation are included in the simulation, except stellar winds. 
Similar simulations with stellar winds shows that their effect on star formation are not completely negligible but modify it only by a factor of two \citep{gattoSILCCProjectIII2017}, and thus cannot explain the discrepancies we observe.
Second, our FUV prescription (uniform heating proportional to the SFR) overestimates the heating because both absorption and the propagation delay are not well taken into account.
Third, additional feedback effects strong enough to reduce star formation to the expected level for  $\Sigma_{\text{gas}, 0} > 25 \Msun\cdot\text{pc}^{-2}$ would probably generate a too weak SFR for simulations with $\Sigma_{\text{gas}, 0} < 20 \Msun\cdot\text{pc}^{-2}$ which are already close to the observed SFR.
Finally, Figure \ref{fig:SK}b shows that the expected turbulent power from stellar feedback is well below what is needed to quench star formation efficiently for high-redshift galaxies.
The inefficiency of stellar feedback to quench star formation in gas-rich galaxies suggests that another phenomenon is likely at play.

\newpage

\section{Effects of turbulence injection}\label{sec:turb}

\begin{figure*}[ht]
    \centering
     \includegraphics[width=0.45\textwidth]{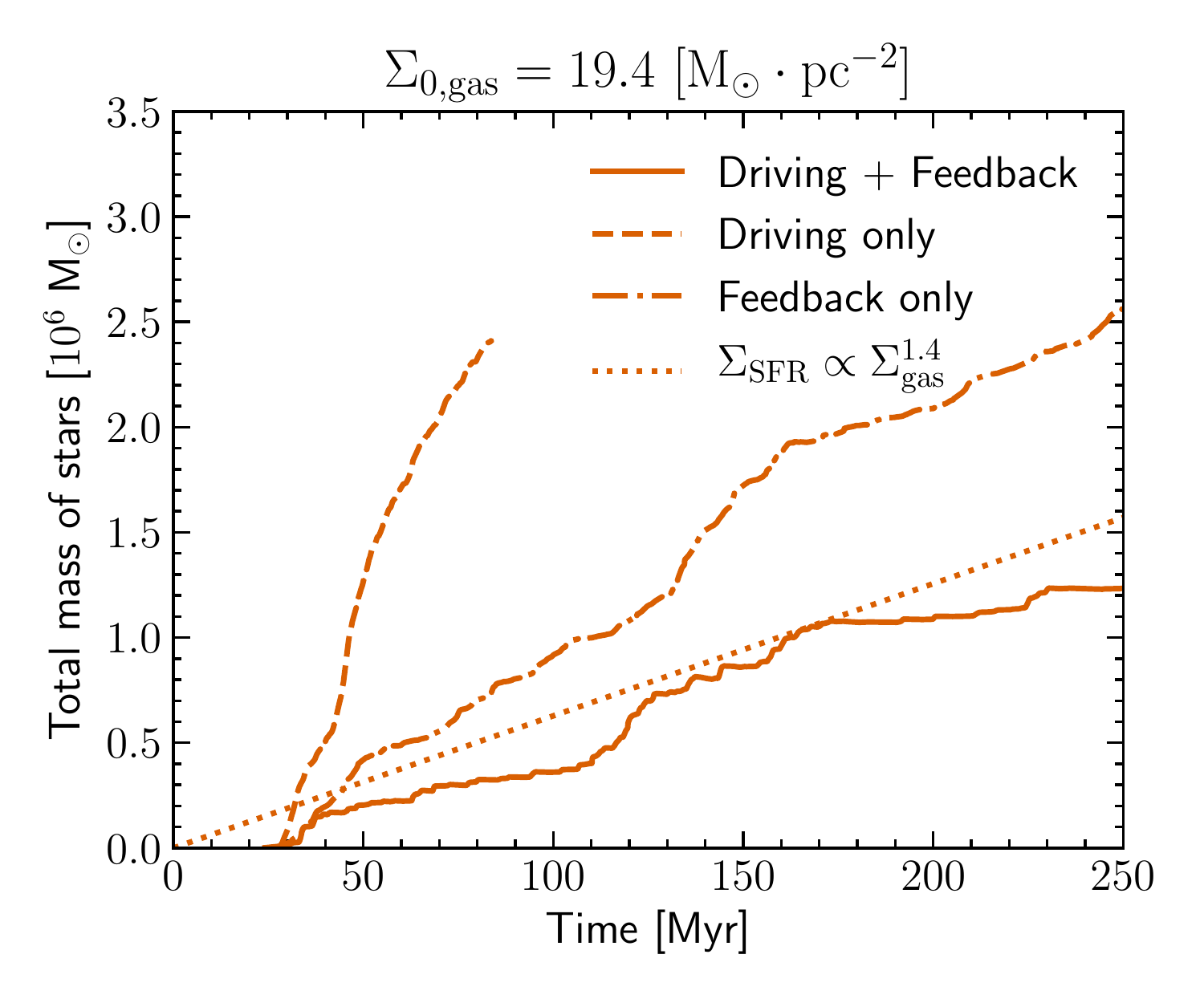}
    \includegraphics[width=0.45\textwidth]{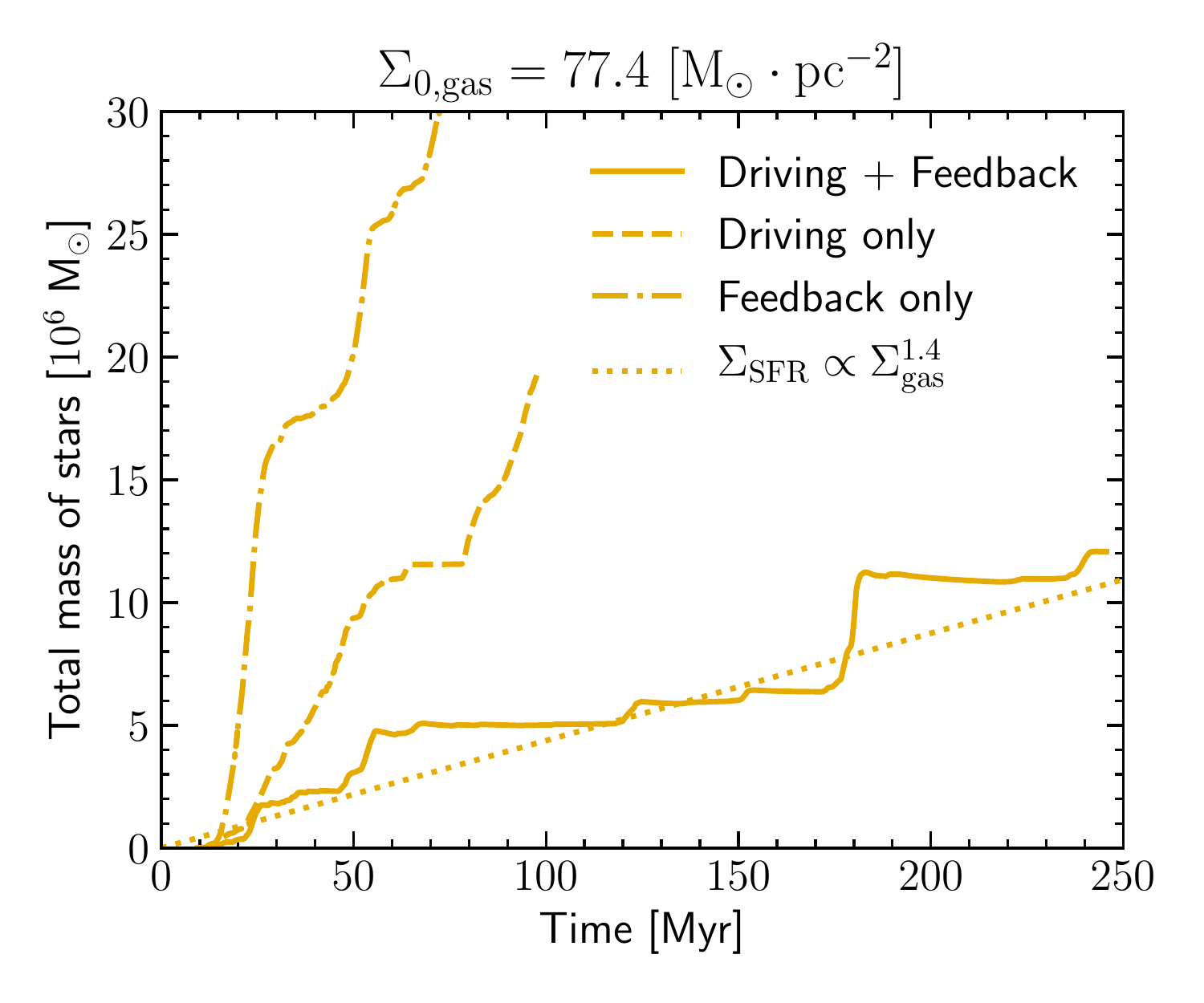}
    \caption{Stellar mass, with and without feedback and turbulence. Feedback and turbulence are needed to quench star formation efficiently.}\label{fig:withoutSF}
\end{figure*}

In the previous section we have shown that a pure stellar feedback was not strong enough to quench star formation efficiently in galaxies with high column density. 
Figure \ref{fig:mass} features the mass accreted by the sinks for several values of the initial gas column density $\Sigma_{0, \mathrm{gas}}$ with a turbulent forcing (with dominant solenoidal modes). 
We tested two scalings for the injected energy, $P_{\mathrm{inj}}\propto\Sigma^{2.5}$ and $P_{\mathrm{inj}}\propto\Sigma^{3.8}$. 
In both sets of simulations, the stellar mass has been reduced from the pure feedback model, and more powerful driving is more efficient at reducing star formation. 
The \textsc{turb3.8} group has stellar mass curve compatible with a SFR matching the SK law.

Indeed, in Figure \ref{fig:SK}a the star formation law derived from this group has an index $N= 1.5$, very close to the $N = 1.4$ of the SK relation. Therefore, large-scale turbulent driving enables to reproduce a formation law close to the SK law when pure stellar feedback cannot.

Turbulent driving has a considerable influence on the shape of the galactic disk, as can be seen in Figure~\ref{fig:coldens} representing the face-on and edge-on column density map of gas with and without turbulent driving.
Pure feedback simulations show a lot of small-scale structures and clumps, and a lot of gas is blasted out of the disk plane by SNe.
When turbulent driving is applied, the gas tends to organize within huge filaments, with fewer and bigger clumps. A significant bulk motion is triggered. The effects of turbulent driving are also clearly visible on the density probability distribution function (PDF) and on the density profile in Figure \ref{fig:density}, in the Appendix. When applied, turbulent driving increases the fraction of gas within low-density regions and can move the position of the disk plane. 
In all cases the scale height of the disk increases for higher value of the column density as the strength of stellar feedback or turbulent driving also increase, but a disk structure is still clearly apparent.
More energetic turbulent driving (or 3D turbulent driving) completely destroys the disks, which sets a limit on the turbulent energy that can be injected. \added{The driving being bidimensional and parallel to the galactic plane, it generates strongly anisotropic velocity dispersion (Figure \ref{fig:sigma}, in the Appendix). The effect increases with the column density. For high-$z$ galaxies, the velocity dispersion alongside the galactic plane $\sigma_{2D} = \sqrt{\sigma_x^2 + \sigma_y^2} / \sqrt{2}$ is 10 times higher than the vertical velocity dispersion $\sigma_z$. By comparison, the velocity dispersion in pure feedback simulations is almost isotropic.}

\section{Is Stellar Feedback Needed at All?}\label{sec:full_turb}

Previous studies \citep{bournaudISMPropertiesHydrodynamic2010,renaudStarFormationLaws2012,krumholzUnifiedModelGalactic2018,hopkinsSelfregulatedStarFormation2011} suggest that both large-scale turbulence and stellar feedback are needed to match observations. 
\cite{blockTwocomponentPowerLaw2010} argue that stellar feedback is crucial to inject energy back to large scales. 
With our setup, we can carry out a simple experiment to see if stellar feedback is necessary to quench star formation.
To investigate this, we rerun two simulations of the \textsc{turb3.8} group, namely those with  $n_0 = 1.5, f_{\mathrm{rms}} = 2\E{4}$ and $n_0 = 6, f_{\mathrm{rms}} = 2\E{5}$, with stellar feedback off (we switch off H~II regions and SNe, and FUV heating is kept constant at solar neighborhood level), so that only the turbulent driving quenches star formation.
On Figure \ref{fig:withoutSF}, we can see that in such a configuration the SFR is higher than the one given by the Kennicutt law. For low gas column density, it is even higher than the one that we obtain with stellar feedback only. 
Thus, it appears that stellar feedback and large-scale turbulence are complementary to quench star formation, and that the relative importance of stellar feedback diminishes as the gas column density increase. This result is in good agreement with the conclusion reached from global galactic simulations \citep{bournaudISMPropertiesHydrodynamic2010}.

\section{Conclusion}\label{sec:conc}

We have presented simulations of kiloparsec cube regions of galaxies with and without stellar feedback and with and without turbulent driving (Table \ref{tbl:simu}, figures \ref{fig:coldens},\ref{fig:withoutSF}). 
The simulated galaxies have a gas column density between $12.9$ and $155 \Msun\cdot\mathrm{pc}^{-2}$.
We reported the SFR in these simulations as function of the gas column density (Figure \ref{fig:mass}) and compared the obtained star formation law with the SK law (Figure \ref{fig:SK}a).
Then we compared the power injected by the turbulent driving needed to reproduce the SK law with estimates of the turbulent power released by large-scale motions and stellar feedback (Figure \ref{fig:SK}b). \added{The effect of the turbulent driving on the velocity dispersion (Figure \ref{fig:sigma}) and the distribution of the gas (Figure \ref{fig:coldens} and \ref{fig:density}) were also studied.} Our main findings are as follows.
\begin{enumerate}
    \item Stellar feedback is able to explain the \added{averaged} SFR in Milky Way--like galaxies. 
    \item In high-redshift galaxies with high gas column densities, stellar feedback alone is too weak to quench star formation to a level consistent with the SK law: the obtained star formation law for the studied range of gas column densities is too steep compared to the SK law. 
    \item The addition of a mainly solenoidal large-scale \added{bidimensional} turbulent driving with a power injection $\overline{P_{\mathrm{inj}}}~\propto~\Sigma^{3.8}$ reduces considerably the SFR. The star formation obtained has an index $N = 1.5$, which is close to the observed SK relation.
    \item The injected power is consistent with the power needed to maintain the disk at marginal stability (with a Toomre $Q \approx 1)$, which scales as $P_{\mathrm{LS}}~\propto~\Sigma^{4}$.
    \item \added{The resulting velocity dispersion is strongly anisotropic. The velocity dispersion parallel to the disk plane $\sigma_{\mathrm{2D}}$ can be up to 10 times higher than the vertical velocity $\sigma_z$.}
    \item Stellar feedback remains necessary, but its importance decreases as the gas column density increases.
\end{enumerate}

Large-scale turbulent driving is therefore necessary when studying star formation in kpc-sized regions of  galaxies, especially when the gas fraction is high.
A key question that arises is what is the exact nature and origin of the turbulence that needs to be injected.

\acknowledgments

\added{We thank the referee for their comments that helped improve the article, and our colleagues for insightful discussions.}
This work was granted access to HPC resources from the TGCC on the Joliot Curie supercomputer under the allocation GENCI A0070407023.

%

\vspace{5mm}


\software{\textsc{Ramses} \citep{teyssierCosmologicalHydrodynamicsAdaptive2002},
          \textsc{Pymses} (Labadens, Chapon, Guillet)}




\appendix

\section{Power injected by the turbulent driving and by the feedback}

\begin{figure*}[h!]
    \centering
     \includegraphics[height=0.27\textheight]{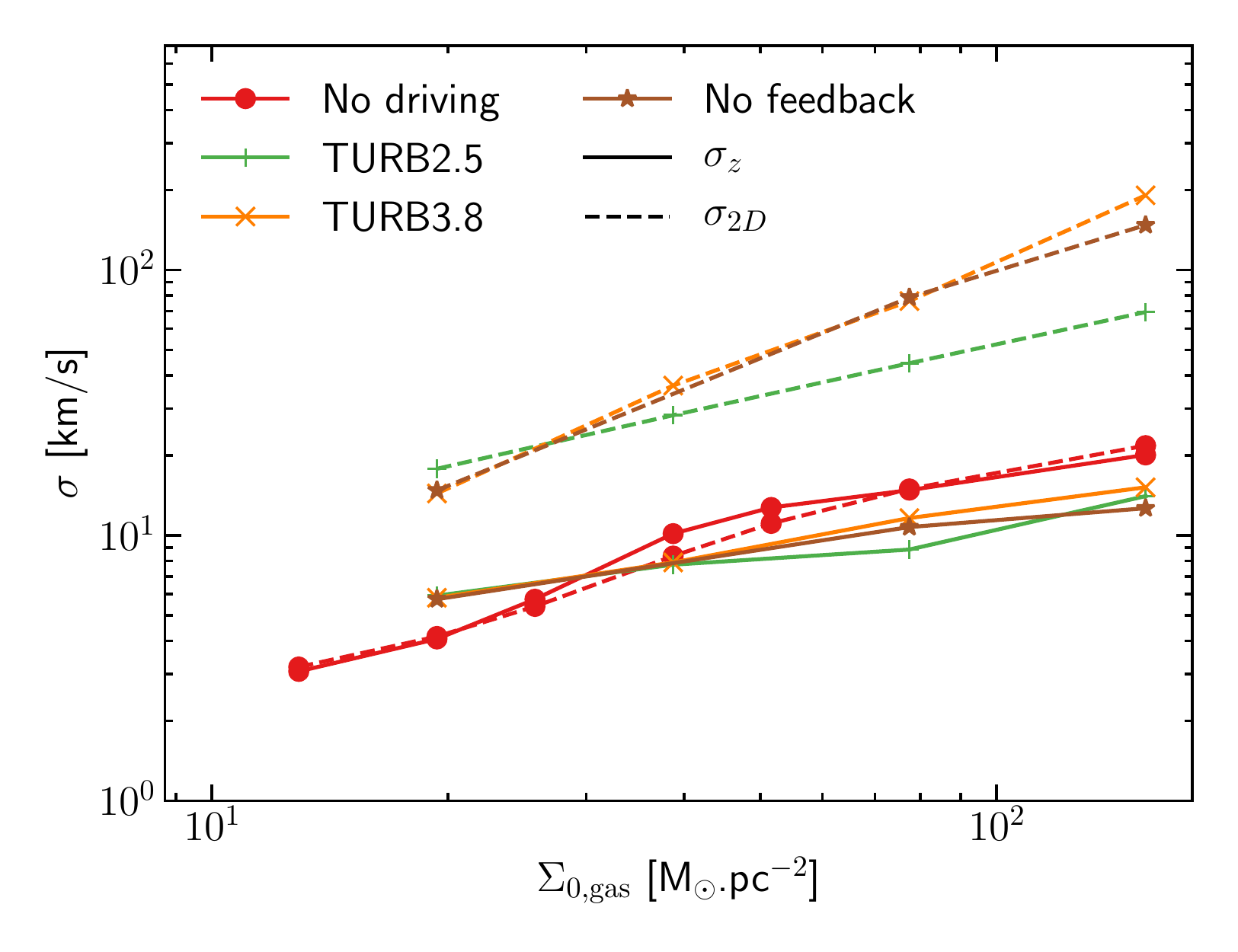}
    \includegraphics[height=0.27\textheight]{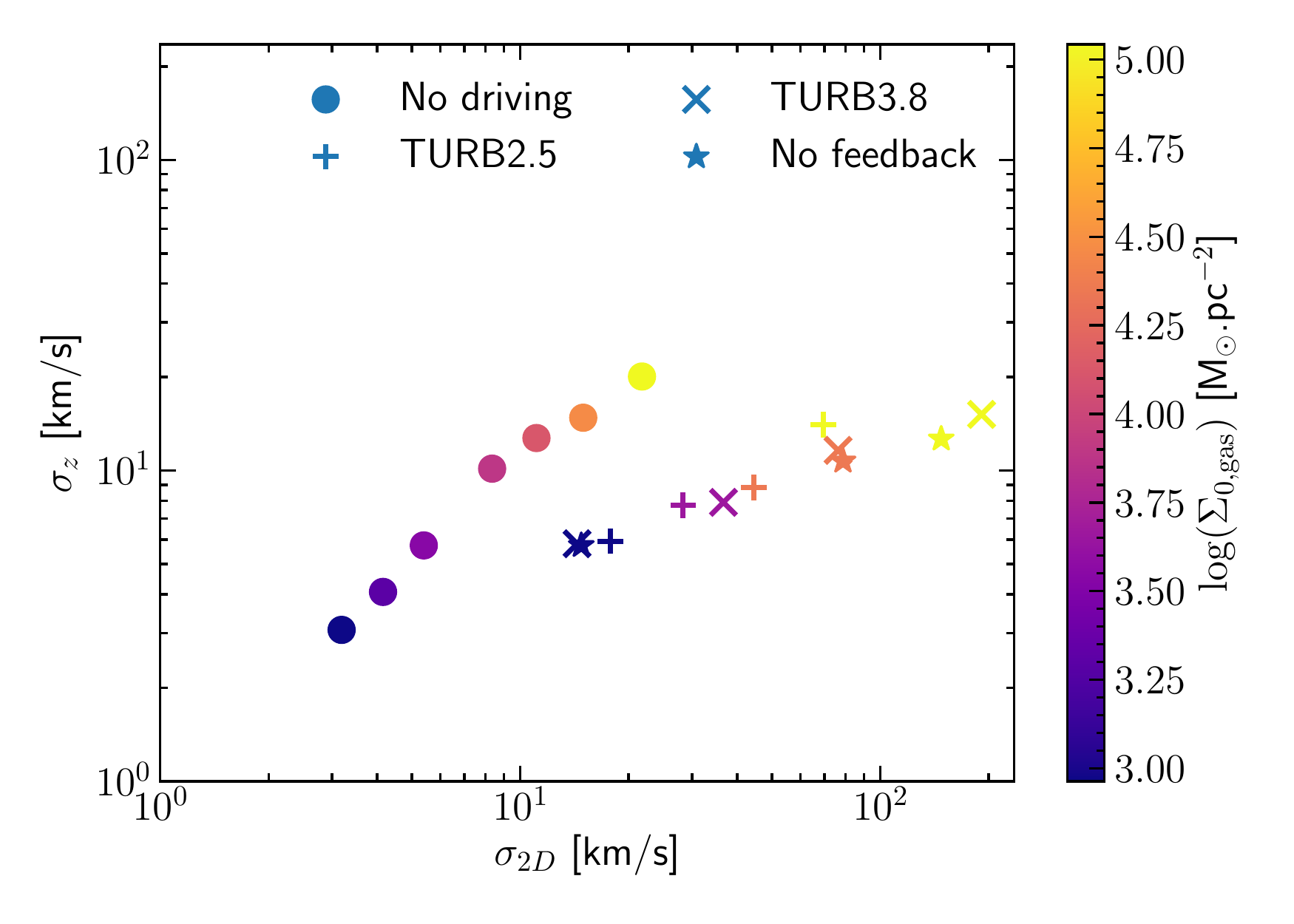}
    \caption{\added{Velocity dispersion measured in the simulations, where $\sigma_{2D} = \sqrt{\sigma_x^2 + \sigma_y^2} / \sqrt{2}$. The simulations with high 2D turbulent driving show a high anisotropy, while simulations without driving are almost isotropic.}}\label{fig:sigma}
\end{figure*}

\label{sec:pinj_abs}
The section \ref{subsec:estimation} provides an estimation on how the power injected via turbulence scales with column density. 
We can go further and estimate what is the absolute value of power injected, and compare it to the value used for the \textsc{turb3.8} group of simulation that best reproduce the SK law and to the power injected by stellar feedback (Figure~\ref{fig:SK}b). To get a relevant value, we must take into account the stellar contribution to the Toomre stability criterion. 
The formula for the Toomre parameter when both the gas and the star fluid are near instability is rather complicated, but the following equation is a acceptable approximation \citep{wangGravitationalInstabilityDisk1994,romeoEffectiveStabilityParameter2011,romeoSimpleAccurateApproximation2013,} 
\begin{equation}
    \frac{1}{Q} = \frac{1}{Q_g} + \frac{1}{Q_\star}
\end{equation}
with
\begin{equation}
\label{eq:Qg}
  Q_g = \frac{\sigma_g \kappa}{\pi \Sigma_g G}\quad \mathrm{and}\quad Q_\star = \frac{\sigma_\star \kappa}{\pi \Sigma_\star G}
\end{equation}

The stability criterion is still $Q \approx 1$. For high-redshift galaxies, $\Sigma_g \approx \Sigma_\star$ \citep{daddiVeryHighGas2010,genzelStudyGasstarFormation2010} and $\sigma_g \approx \sigma_\star$ (as reported by \cite{elmegreenObservationsThickDisks2006} with measurement based on the thickness of edge-on stellar disks). In $z = 0$ Milky Way--like galaxies, $\Sigma_g \approx 0.1\ \Sigma_\star$ \citep{deblokHighResolutionRotationCurves2008} and $\sigma_g \approx 0.1\ \sigma_\star$ \cite[][and references therein]{falcon-barrosoStellarKinematicsHubble2017,hennebelleTurbulentMolecularClouds2012}. In both cases, $Q_g \approx Q_\star$ and then

\begin{equation}
\label{eq:Qg2}
  Q_g \approx 2
\end{equation}

Using equations \ref{eq:injection} and \ref{eq:Qg}  we get
\begin{equation}
\label{eq:pinj_abs}
    P_{\mathrm{LS}} = \Sigma_g L^2 \cdot 2\frac{\sigma_g^3}{L} =  \frac{2 L Q_g^3 \pi^3 G^3}{\kappa^3}  \Sigma_g^4
\end{equation}
where we take $l = L / 2$ as typical injection scale (see section \ref{subsec:turb_inj}), with $L = 1\  \mathrm{kpc}$ the length of one side the box. We take solar neighborhood value for the epicyclic frequency $\kappa$:
\begin{equation}
    \kappa \approx \sqrt{2} \Omega \approx \sqrt{2} \frac{v}{R}
\end{equation}
with $v = 220\ \mathrm{km}\cdot \mathrm{s}^{-1}$ and $R = 8\ \mathrm{kpc}$. As a result,

\begin{equation}
    P_{\mathrm{LS}} \approx 4.3 \E{29} \left(\frac{\Sigma_g}{10 \Msun\cdot\mathrm{pc}^{-2}} \right)^4\ \mathrm{W}
\end{equation}

This value is probably a lower bound because the values of the velocity dispersion reported in the observations are usually derived under the assumption of isotropy.
However, the velocity dispersion at the scales that we look at is dominated by the 2D velocity dispersion within the disk \replaced{, and thus may be underestimated}{(Figure \ref{fig:sigma})}.
\added{The shaded blue region in Figure \ref{fig:SK}b show the range of values of $P_{\mathrm{LS}}$ if this underestimation was of a factor of one to two. }

Figure \ref{fig:SK}b emphasizes another important fact: it is completely unlikely that our turbulent driving mimick the effect of stellar feedback-driven turbulence. 
Indeed, the energy injected under the form of turbulence by the stellar feedback scales as the SFR, that is $P_{\mathrm{feedback}} \propto \Sigma_{\mathrm{SFR}}  \propto \Sigma_{g}^{1.4}$, which is not compatible with the relation $\overline{P_{\mathrm{inj}}} \propto \Sigma_{g}^{3.8}$ needed to reproduce the SK law. On Figure \ref{fig:SK}b, we illustrate this with an estimation of the energy injected by the dominant feedback mechanism, SNe.
There is approximately one SN each time $100 \Msun$ of stellar mass is created. It releases $10^{51}$ erg into the interstellar medium. \cite{iffrigMutualInfluenceSupernovae2015} and \cite{martizziSupernovaFeedbackLocal2016} have shown that at these scales, only a fraction of a few percent of this energy is converted into turbulence. \replaced{We take $5 \%$ as an upper bound}{We retain values between $1 \%$ and $5 \%$ as reasonable (red shaded region in Figure \ref{fig:SK}b)}.
The \added{upper bound for the} turbulent power injected by the SN  is then
\begin{equation}
    P_{\mathrm{SN}} \approx 4.0 \E{30} \left(\frac{\Sigma_g}{10 \Msun\cdot\mathrm{pc}^{-2}} \right)^{1.4}\ \mathrm{W}
\end{equation}
It is clearly not sufficient for high-redshift galaxies, but dominates over the power $P_{\mathrm{LS}}$ as estimated in Equation \ref{eq:pinj_abs} for Milky way--like galaxies. 
This is coherent with our result that stellar feedback alone is sufficient in such galaxies.

\section{Effect of turbulence on density distribution}

\begin{figure*}[ht]
    \centering
    \includegraphics[width=\textwidth]{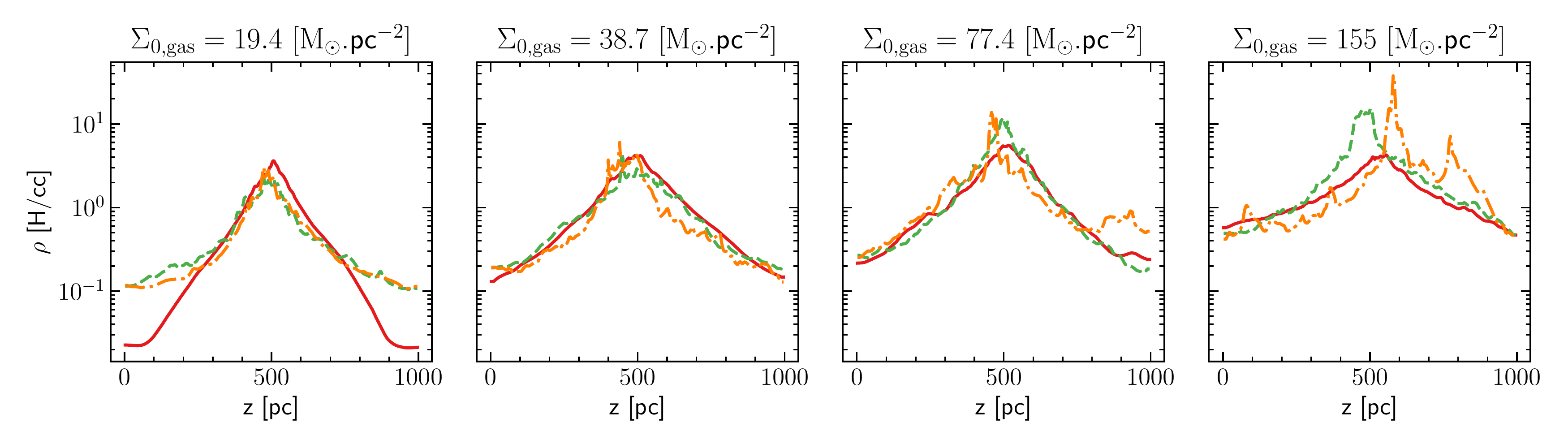}\\
    \includegraphics[width=\textwidth]{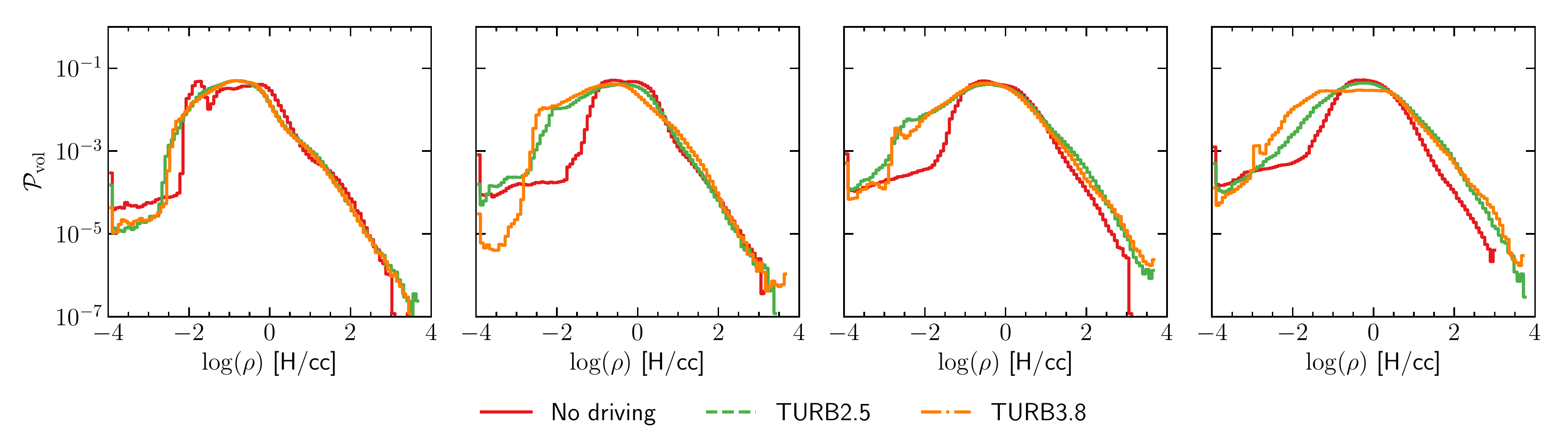}
    \caption{Averaged density profile, top, and density volumic probability distribution function (PDF), bottom. All figures are made from snapshots taken at $t \approx 60$ Myr. \added{There is less dense gas in the simulations with high initial colunm density ($\Sigma_{0,\mathrm{gas}} \geq 77 \Msun.\mathrm{pc}^{-2}$) without driving because most it has been accreted by the sinks.} }\label{fig:density}
\end{figure*}

The Figure \ref{fig:density} gives more insight on the effects of the turbulence on the density distribution.
The density profile shows that all simulations feature a stratified gas distribution, and that the profile is less steep when the gas column density or the turbulence forcing increase.
Strong turbulence (for $\Sigma_{0,\mathrm{gas}} = 155$) can trigger huge bulk motion that can move the position of the disk plane. Stronger turbulence can even disrupt the disk.
The turbulent driving redistributes the gas and widens the gas PDF, increasing the fraction of gas in low-density regions, diminishing the gas available for star formation.

\added{The simulations without driving convert a subsequent fraction of the gas into star because of the high SFR. At 60~Myr, 39 \% and 58 \%, respectively, of the total initial mass of gas in the box was accreted by the sinks for the $\Sigma_{0, \mathrm{gas}} = 77.4 \Msun$.pc$^{-2}$ and $\Sigma_{0, \mathrm{gas}} = 155 \Msun$.pc$^{-2}$ simulations without driving. 
This mass is took from the densest regions of the box, and as a consequence there is less dense gas remaining in the box. By contrast, for the same simulations with driving (group \textsc{turb3.8}) only about 6\% of gas has been accreted at 60 Myr. 
This explains why the simulations without driving have less dense gas that the corresponding simulations with driving.
}


\newpage

\begin{thebibliography}{}
\expandafter\ifx\csname natexlab\endcsname\relax\def\natexlab#1{#1}\fi
\providecommand{\url}[1]{\href{#1}{#1}}

\bibitem[{Bleuler \& Teyssier(2014)}]{bleulerMoreRealisticSink2014}
Bleuler, A., \& Teyssier, R. 2014, Monthly Notices of the Royal Astronomical
  Society, 445, 4015.
\newblock \url{http://adsabs.harvard.edu/abs/2014MNRAS.445.4015B}

\bibitem[{Block {et~al.}(2010)Block, Puerari, Elmegreen, \&
  Bournaud}]{blockTwocomponentPowerLaw2010}
Block, D.~L., Puerari, I., Elmegreen, B.~G., \& Bournaud, F. 2010, The
  Astrophysical Journal Letters, 718, L1.
\newblock \url{http://adsabs.harvard.edu/abs/2010ApJ...718L...1B}

\bibitem[{Bournaud(2014)}]{bournaudStarFormationLaws2014}
Bournaud, F. 2014, ASP Conference Series, 486, 101.
\newblock \url{http://adsabs.harvard.edu/abs/2014ASPC..486..101B}

\bibitem[{Bournaud {et~al.}(2010)Bournaud, Elmegreen, Teyssier, Block, \&
  Puerari}]{bournaudISMPropertiesHydrodynamic2010}
Bournaud, F., Elmegreen, B.~G., Teyssier, R., Block, D.~L., \& Puerari, I.
  2010, <pre>\mnras</pre>, 409, 1088.
\newblock \url{http://dx.doi.org/10.1111/j.1365-2966.2010.17370.x}

\bibitem[{Colling {et~al.}(2018)Colling, Hennebelle, Geen, Iffrig, \&
  Bournaud}]{collingImpactGalacticShear2018}
Colling, C., Hennebelle, P., Geen, S., Iffrig, O., \& Bournaud, F. 2018, \aap,
  620, A21

\bibitem[{Daddi {et~al.}(2010)Daddi, Bournaud, Walter, Dannerbauer, Carilli,
  Dickinson, Elbaz, Morrison, Riechers, Onodera, Salmi, Krips, \&
  Stern}]{daddiVeryHighGas2010}
Daddi, E., Bournaud, F., Walter, F., {et~al.} 2010, The Astrophysical Journal,
  713, 686.
\newblock \url{http://adsabs.harvard.edu/abs/2010ApJ...713..686D}

\bibitem[{{de Blok} {et~al.}(2008){de Blok}, Walter, Brinks, Trachternach, Oh,
  \& Kennicutt}]{deblokHighResolutionRotationCurves2008}
{de Blok}, W. J.~G., Walter, F., Brinks, E., {et~al.} 2008, The Astronomical
  Journal, 136, 2648.
\newblock \url{http://adsabs.harvard.edu/abs/2008AJ....136.2648D}

\bibitem[{Dekel {et~al.}(2009)Dekel, Sari, \&
  Ceverino}]{dekelFormationMassiveGalaxies2009}
Dekel, A., Sari, R., \& Ceverino, D. 2009, The Astrophysical Journal, 703, 785.
\newblock \url{http://adsabs.harvard.edu/abs/2009ApJ...703..785D}

\bibitem[{Dubois \& Teyssier(2008)}]{duboisOnsetGalacticWinds2008}
Dubois, Y., \& Teyssier, R. 2008, Astronomy and Astrophysics, 477, 79.
\newblock \url{http://adsabs.harvard.edu/abs/2008A\%26A...477...79D}

\bibitem[{Elmegreen \& Elmegreen(2006)}]{elmegreenObservationsThickDisks2006}
Elmegreen, B.~G., \& Elmegreen, D.~M. 2006, The Astrophysical Journal, 650,
  644.
\newblock \url{http://adsabs.harvard.edu/abs/2006ApJ...650..644E}

\bibitem[{Emsellem {et~al.}(2015)Emsellem, Renaud, Bournaud, Elmegreen, Combes,
  \& Gabor}]{emsellemInterplayGalacticBar2015}
Emsellem, E., Renaud, F., Bournaud, F., {et~al.} 2015, Monthly Notices of the
  Royal Astronomical Society, 446, 2468.
\newblock \url{http://adsabs.harvard.edu/abs/2015MNRAS.446.2468E}

\bibitem[{Eswaran \& Pope(1988)}]{eswaranExaminationForcingDirect1988}
Eswaran, V., \& Pope, S.~B. 1988, Computers and Fluids, 16, 257

\bibitem[{{Falc{\'o}n-Barroso} {et~al.}(2017){Falc{\'o}n-Barroso}, Lyubenova,
  van~de Ven, {Mendez-Abreu}, Aguerri, {Garc{\'i}a-Lorenzo}, Bekerait{\'e},
  S{\'a}nchez, Husemann, {Garc{\'i}a-Benito}, Mast, Walcher, Zibetti,
  {Barrera-Ballesteros}, Galbany, {S{\'a}nchez-Bl{\'a}zquez}, Singh, van~den
  Bosch, Wild, Zhu, {Bland-Hawthorn}, Fernandes, de~{Lorenzo-C{\'a}ceres},
  Gallazzi, Delgado, Marino, M{\'a}rquez, P{\'e}rez, P{\'e}rez, Roth,
  {Rosales-Ortega}, {Ruiz-Lara}, Wisotzki, \&
  Ziegler}]{falcon-barrosoStellarKinematicsHubble2017}
{Falc{\'o}n-Barroso}, J., Lyubenova, M., van~de Ven, G., {et~al.} 2017,
  Astronomy \& Astrophysics, 597, A48.
\newblock
  \url{https://www.aanda.org/articles/aa/abs/2017/01/aa28625-16/aa28625-16.html}

\bibitem[{Federrath {et~al.}(2010)Federrath, Roman-Duval, Klessen, Schmidt, \&
  Mac~Low}]{federrathComparingStatisticsInterstellar2010}
Federrath, C., Roman-Duval, J., Klessen, R.~S., Schmidt, W., \& Mac~Low, M.~M.
  2010, \aap, 512, A81.
\newblock \url{https://ui.adsabs.harvard.edu/abs/2010A\&A...512A..81F}

\bibitem[{Fromang {et~al.}(2006)Fromang, Hennebelle, \&
  Teyssier}]{fromangHighOrderGodunov2006}
Fromang, S., Hennebelle, P., \& Teyssier, R. 2006, \aap, 457, 371.
\newblock \url{http://dx.doi.org/10.1051/0004-6361:20065371}

\bibitem[{Gatto {et~al.}(2017)Gatto, Walch, Naab, Girichidis, W{\"u}nsch,
  Glover, Klessen, Clark, Peters, Derigs, Baczynski, \&
  Puls}]{gattoSILCCProjectIII2017}
Gatto, A., Walch, S., Naab, T., {et~al.} 2017, Monthly Notices of the Royal
  Astronomical Society, 466, 1903.
\newblock \url{http://adsabs.harvard.edu/abs/2017MNRAS.466.1903G}

\bibitem[{Genzel {et~al.}(2008)Genzel, Burkert, Bouché, Cresci,
  Förster~Schreiber, Shapley, Shapiro, Tacconi, Buschkamp, Cimatti, Daddi,
  Davies, Eisenhauer, Erb, Genel, Gerhard, Hicks, Lutz, Naab, Ott, Rabien,
  Renzini, Steidel, Sternberg, \& Lilly}]{genzelRingsBulgesEvidence2008}
Genzel, R., Burkert, A., Bouché, N., {et~al.} 2008, \apj, 687, 59

\bibitem[{Genzel {et~al.}(2010)Genzel, Tacconi, {Gracia-Carpio}, Sternberg,
  Cooper, Shapiro, Bolatto, Bouch{\'e}, Bournaud, Burkert, Combes, Comerford,
  Cox, Davis, Schreiber, {Garcia-Burillo}, Lutz, Naab, Neri, Omont, Shapley, \&
  Weiner}]{genzelStudyGasstarFormation2010}
Genzel, R., Tacconi, L.~J., {Gracia-Carpio}, J., {et~al.} 2010, Monthly Notices
  of the Royal Astronomical Society, 407, 2091.
\newblock \url{http://adsabs.harvard.edu/abs/2010MNRAS.407.2091G}

\bibitem[{Hennebelle \&
  Falgarone(2012)}]{hennebelleTurbulentMolecularClouds2012}
Hennebelle, P., \& Falgarone, E. 2012, Astronomy and Astrophysics Review, 20,
  55.
\newblock \url{http://adsabs.harvard.edu/abs/2012A\%26ARv..20...55H}

\bibitem[{Hopkins {et~al.}(2011)Hopkins, Quataert, \&
  Murray}]{hopkinsSelfregulatedStarFormation2011}
Hopkins, P.~F., Quataert, E., \& Murray, N. 2011, Monthly Notices of the Royal
  Astronomical Society, 417, 950.
\newblock \url{http://adsabs.harvard.edu/abs/2011MNRAS.417..950H}

\bibitem[{Iffrig \& Hennebelle(2015)}]{iffrigMutualInfluenceSupernovae2015}
Iffrig, O., \& Hennebelle, P. 2015, Astronomy and Astrophysics, 576, A95.
\newblock \url{http://adsabs.harvard.edu/abs/2015A\%26A...576A..95I}

\bibitem[{Iffrig \&
  Hennebelle(2017)}]{iffrigStructureDistributionTurbulence2017}
---. 2017, \aap, 604, A70

\bibitem[{Jin {et~al.}(2017)Jin, Salim, Federrath, Tasker, Habe, \&
  Kainulainen}]{jinEffectiveTurbulenceDriving2017}
Jin, K., Salim, D.~M., Federrath, C., {et~al.} 2017, Monthly Notices of the
  Royal Astronomical Society, 469, 383.
\newblock \url{http://adsabs.harvard.edu/abs/2017MNRAS.469..383J}

\bibitem[{Kennicutt \& Evans(2012)}]{kennicuttStarFormationMilky2012}
Kennicutt, R.~C., \& Evans, N.~J. 2012, Annual Review of Astronomy and
  Astrophysics, 50, 531.
\newblock \url{http://adsabs.harvard.edu/abs/2012ARA\%26A..50..531K}

\bibitem[{Kennicutt(1998)}]{kennicuttjr.GlobalSchmidtLaw1998}
Kennicutt, Jr., R.~C. 1998, \apj, 498, 541.
\newblock \url{http://dx.doi.org/10.1086/305588}

\bibitem[{Kim \& Ostriker(2017)}]{kimThreephaseInterstellarMedium2017}
Kim, C.-G., \& Ostriker, E.~C. 2017, The Astrophysical Journal, 846, 133.
\newblock \url{http://adsabs.harvard.edu/abs/2017ApJ...846..133K}

\bibitem[{Kolmogorov(1941)}]{kolmogorovLocalStructureTurbulence1941}
Kolmogorov, A. 1941, Akademiia Nauk SSSR Doklady, 30, 301

\bibitem[{Krumholz \&
  Burkhart(2016)}]{krumholzTurbulenceInterstellarMedium2016}
Krumholz, M.~R., \& Burkhart, B. 2016, Monthly Notices of the Royal
  Astronomical Society, 458, 1671.
\newblock \url{http://adsabs.harvard.edu/abs/2016MNRAS.458.1671K}

\bibitem[{Krumholz {et~al.}(2018)Krumholz, Burkhart, Forbes, \&
  Crocker}]{krumholzUnifiedModelGalactic2018}
Krumholz, M.~R., Burkhart, B., Forbes, J.~C., \& Crocker, R.~M. 2018, Monthly
  Notices of the Royal Astronomical Society, 477, 2716.
\newblock \url{https://academic.oup.com/mnras/article/477/2/2716/4962399}

\bibitem[{Martizzi {et~al.}(2016)Martizzi, Fielding, {Faucher-Gigu{\`e}re}, \&
  Quataert}]{martizziSupernovaFeedbackLocal2016}
Martizzi, D., Fielding, D., {Faucher-Gigu{\`e}re}, C.-A., \& Quataert, E. 2016,
  Monthly Notices of the Royal Astronomical Society, 459, 2311.
\newblock \url{http://adsabs.harvard.edu/abs/2016MNRAS.459.2311M}

\bibitem[Meidt et al.(2020)]{meidtModelOnsetSelfgravitation2020} Meidt, S.~E., Glover, S.~C.~O., Kruijssen, J.~M.~D., et al.\ 2020, \apj, 892, 73

\bibitem[{Ostriker {et~al.}(2010)Ostriker, McKee, \&
  Leroy}]{ostrikerRegulationStarFormation2010}
Ostriker, E.~C., McKee, C.~F., \& Leroy, A.~K. 2010, \apj, 721, 975

\bibitem[{Padoan {et~al.}(2016)Padoan, Pan, Haugb{\o}lle, \&
  Nordlund}]{padoanSupernovaDrivingOrigin2016}
Padoan, P., Pan, L., Haugb{\o}lle, T., \& Nordlund, {\AA}. 2016, The
  Astrophysical Journal, 822, 11.
\newblock \url{http://adsabs.harvard.edu/abs/2016ApJ...822...11P}

\bibitem[{Renaud {et~al.}(2012)Renaud, Kraljic, \&
  Bournaud}]{renaudStarFormationLaws2012}
Renaud, F., Kraljic, K., \& Bournaud, F. 2012, The Astrophysical Journal
  Letters, 760, L16.
\newblock \url{http://adsabs.harvard.edu/abs/2012ApJ...760L..16R}

\bibitem[{Romeo \& Falstad(2013)}]{romeoSimpleAccurateApproximation2013}
Romeo, A.~B., \& Falstad, N. 2013, Monthly Notices of the Royal Astronomical
  Society, 433, 1389.
\newblock \url{http://adsabs.harvard.edu/abs/2013MNRAS.433.1389R}

\bibitem[{Romeo \& Wiegert(2011)}]{romeoEffectiveStabilityParameter2011}
Romeo, A.~B., \& Wiegert, J. 2011, Monthly Notices of the Royal Astronomical
  Society, 416, 1191.
\newblock \url{http://adsabs.harvard.edu/abs/2011MNRAS.416.1191R}

\bibitem[{Schmidt {et~al.}(2009)Schmidt, Federrath, Hupp, Kern, \&
  Niemeyer}]{schmidtNumericalSimulationsCompressively2009}
Schmidt, W., Federrath, C., Hupp, M., Kern, S., \& Niemeyer, J.~C. 2009, \aap,
  494, 127.
\newblock \url{http://dx.doi.org/10.1051/0004-6361:200809967}

\bibitem[{Schmidt {et~al.}(2006)Schmidt, Hillebrandt, \&
  Niemeyer}]{schmidtNumericalDissipationBottleneck2006}
Schmidt, W., Hillebrandt, W., \& Niemeyer, J.~C. 2006, Computers \& Fluids, 35,
  353.
\newblock
  \url{http://www.sciencedirect.com/science/article/pii/S0045793005000563}

\bibitem[{Teyssier(2002)}]{teyssierCosmologicalHydrodynamicsAdaptive2002}
Teyssier, R. 2002, Astronomy and Astrophysics, 385, 337.
\newblock \url{https://ui.adsabs.harvard.edu/abs/2002A\&A...385..337T}

\bibitem[{Walch {et~al.}(2015)Walch, Girichidis, Naab, Gatto, Glover,
  W{\"u}nsch, Klessen, Clark, Peters, Derigs, \&
  Baczynski}]{walchSILCCSImulatingLifeCycle2015}
Walch, S., Girichidis, P., Naab, T., {et~al.} 2015, Monthly Notices of the
  Royal Astronomical Society, 454, 238.
\newblock \url{http://adsabs.harvard.edu/abs/2015MNRAS.454..238W}

\bibitem[{Wang \& Silk(1994)}]{wangGravitationalInstabilityDisk1994}
Wang, B., \& Silk, J. 1994, The Astrophysical Journal, 427, 759.
\newblock \url{http://adsabs.harvard.edu/abs/1994ApJ...427..759W}

\end{thebibliography}

\listofchanges

\end{document}